\documentclass[sigconf]{acmart}

\usepackage{booktabs} 

\usepackage[caption=false,font=normalsize]{subfig}
\usepackage{flushend} 
\usepackage[linesnumbered,ruled]{algorithm2e}

\copyrightyear{2018} 
\acmYear{2018} 
\setcopyright{licensedusgovmixed}
\acmConference[DYNAMICS '18]{DYnamic and Novel Advances in Machine Learning and Intelligent Cyber Security Workshop}{December 2018}{San Juan, Puerto Rico, USA}
\acmPrice{15.00}
\acmDOI{10.1145/3306195.3306198}
\acmISBN{978-1-4503-6218-4/18/12}

\begin{document}
\title{Cyber Anomaly Detection Using Graph-node Role-dynamics}

\author{Anthony Palladino}
\orcid{0000-0002-2697-3869}
\affiliation{%
  \institution{Boston Fusion Corp.}
  \streetaddress{70 Westview Street, Suite 100}
  \city{Lexington}
  \state{Massachusetts}
  \postcode{02421}
}
\email{anthony.palladino@bostonfusion.com}

\author{Christopher J. Thissen}
\orcid{0000-0002-8714-7755}
\affiliation{%
  \institution{Boston Fusion Corp.}
  \streetaddress{70 Westview Street, Suite 100}
  \city{Lexington}
  \state{Massachusetts}
  \postcode{02421}
}
\email{christopher.thissen@bostonfusion.com}

\renewcommand{\shortauthors}{A.~Palladino and C.J.~Thissen}

\begin{abstract}
Intrusion detection systems (IDSs) generate valuable knowledge about network security, but an abundance of false alarms and a lack of methods to capture the interdependence among alerts hampers their utility for network defense. Here, we explore a graph-based approach for fusing alerts generated by multiple IDSs (e.g., Snort, OSSEC, and Bro). Our approach generates a weighted graph of alert fields (not network topology) that makes explicit the connections between multiple alerts, IDS systems, and other cyber artifacts. 
We use this multi-modal graph to identify anomalous changes in the alert patterns of a network. To detect the anomalies, we apply the role-dynamics approach, which has successfully identified anomalies in social media, email, and IP communication graphs. In the cyber domain, each node (alert field) in the fused IDS alert graph is assigned a probability distribution across a small set of roles based on that node's features. A cyber attack should trigger IDS alerts and cause changes in the node features, but rather than track every feature for every alert-field node individually, roles provide a succinct, integrated summary of those feature changes. We measure changes in each node's probabilistic role assignment over time, and identify anomalies as deviations from expected roles.

We test our approach using IDS alerts generated from a network of 24~virtual machines (workstations, data and print servers, DHCP and DNS servers), virtual switches, and a virtual server that approximates connections to the internet. The simulation includes three weeks of normal background traffic, as well as  cyber attacks that occur near the end of the simulations. The network includes installations of Snort and OSSEC, which generated alerts throughout the experiment. A NetFlow sensor also captured the network traffic during the simulation. This paper presents a novel approach to multi-modal data fusion and a novel application of role dynamics within the cyber-security domain. Our results show a drastic decrease in the false-positive rate when considering our anomaly indicator instead of the IDS alerts themselves, thereby reducing alarm fatigue and providing a promising avenue for threat intelligence in network defense.

\end{abstract}

%
%



\maketitle

\section{Introduction}

Sophisticated modern cyber attackers, such as Advanced Persistent Threats (APTs), pose a serious threat to critical cyber infrastructure. The term APT generally refers to experienced cyber groups that are directed and supported by governments, corporations, terrorist groups, or other entities motivated by political and economic gain~\cite{hutchins2011intelligence, vukalovic:2015, lemay:2018}. APTs have successfully infiltrated democratic institutions~\cite{lemay:2018}, health insurance companies~\cite{lemay:2018}, and financial institutions (e.g., Equifax~\cite{equifax:2017, equifax:2018}). Specific goals differ and vary over time; prior attacks have exfiltrated sensitive data (e.g., OPM breach) and crippled physical infrastructure (e.g., Stuxnet). Lemay et~al.~\cite{lemay:2018} provide an overview of recent APTs.

Defending against these advanced attacks has proven difficult. APTs leverage social engineering, build advanced cyber-attack tools, and exploit target-specific vulnerabilities~\cite{lemay:2018, RSA:2011, KrekelBakos:2009, sans:2011, damballa:2010, khalil:2015}. Some APTs have altered tactics after their methods have been divulged or countered~\cite{lemay:2018, RSA:2011, KrekelBakos:2009, sans:2011, damballa:2010}. Most prioritize stealth to maintain long-term access and distribute attack steps over months or years~\cite{cole2012advanced} to temporally separate indicators of compromise. 
These aspects make each APT attack unique in most respects, and traditional signature-based detection schemes are mostly ineffective for detecting them. This is confirmed in~\cite{verizon:2010}, which found that in 86\% of cases, evidence about the data breach was recorded in the organization logs, but the detection mechanisms failed to raise security alarms. 

Detecting APTs in networks remains a top research priority~\cite{cole2012advanced}. Detection provides opportunities to: mobilize cyber defense; further study state-of-the-art tactics, techniques, and procedures; attribute an attack to a specific APT or supporting organization; and initiate counter attacks. Although APT methods vary, previous work has outlined abstract phases that characterize the sequence of events in an attack. A widely used example is the intrusion kill chain~\cite{hutchins2011intelligence}: (\emph{i}) reconnaissance, (\emph{ii}) weaponization, (\emph{iii}) delivery, (\emph{iv}) exploitation, (\emph{v}) installation, (\emph{vi}) command and control, and (\emph{vii}) actions on objectives. 

Regardless of the abstraction, attack steps during these phases leave detectable artifacts. These artifacts, though incomplete, provide important information about APTs. Much of our knowledge about APTs results from cyber forensic investigations of these artifacts (e.g.,~\cite{lemay:2018}). The success of cyber forensic analyses (e.g.,~\cite{khalil:2015, grizzly:2016}) suggests that a sufficiently advanced system may be able to provide early detection from these artifacts. A key challenge is developing a method capable of correlating indicators of compromise (IOCs) widely spread in time and modality. 

We present a machine learning approach for detecting attacks from cyber artifacts. Our approach uses unsupervised graph-based anomaly detection~\cite{Akoglu:2015:GBA:2757532.2757629}. Graphs easily fuse multi-modal data and explicitly capture connections between artifacts that are otherwise absent or only implicit in non-graph methods (e.g.,~\cite{julisch:2001}). Note that the graph represents connections between artifacts, and not network topology. Two hosts that frequently communicate may not be directly connected in the graph, just as two hosts that never communicate may be linked by artifacts.

Our unsupervised method learns the artifacts generated by normal network usage and flags anomalies in new data for further inspection. It does not require labeled training data. 
We demonstrate our graph-based anomaly detector by analyzing alert logs generated by network and host intrusion detection systems (IDSs). While APTs may use custom tools or zero-day exploits that circumvent IDSs, it is also not uncommon for APTs, in some phases, to use more widely available malware that does trigger alerts from IDSs~\cite{vukalovic:2015}. A common problem with IDSs is alarm fatigue; attacks may trigger IDS alerts, but so does normal network usage~\cite{julisch:2001}. This allows an APT, even while generating IDS alerts, to remain undetected in a network. Problematically, the same alerts may be triggered by both normal usage and while under an attack, so APT detection requires more than careful IDS tuning. Instead, the challenge is to identify when a set of IDS alerts are triggered by normal activity (false alarms) and when they are triggered by an attack (true alarms).

We test our method using a virtual network that simulates both normal usage and attacks characteristic of the Hurricane Panda and Energetic Bear APTs~\cite{lemay:2018}. Our experiments show that graph feature anomalies are sensitive indicators of APT-like tactics, techniques, and procedures and other types of cyber attacks.

\section{Cyber Artifact Fusion Using Graphs}

Indications of a cyber attack are often split across modalities. A SQL injection attack, for example, may trigger alerts from a network IDS~\cite{alnabulsi:2014}, while adding new admin users may trigger a different, host-based IDS\footnote{https://ossec.github.io/}. Rather than treat these alerts as independent, an artifact graph explicitly captures the links between the alerts. In this example, those links might include overlapping IP addresses if both attacks involved the same machine. 

\begin{figure}[!hb]
\centering
\includegraphics[width=\columnwidth]{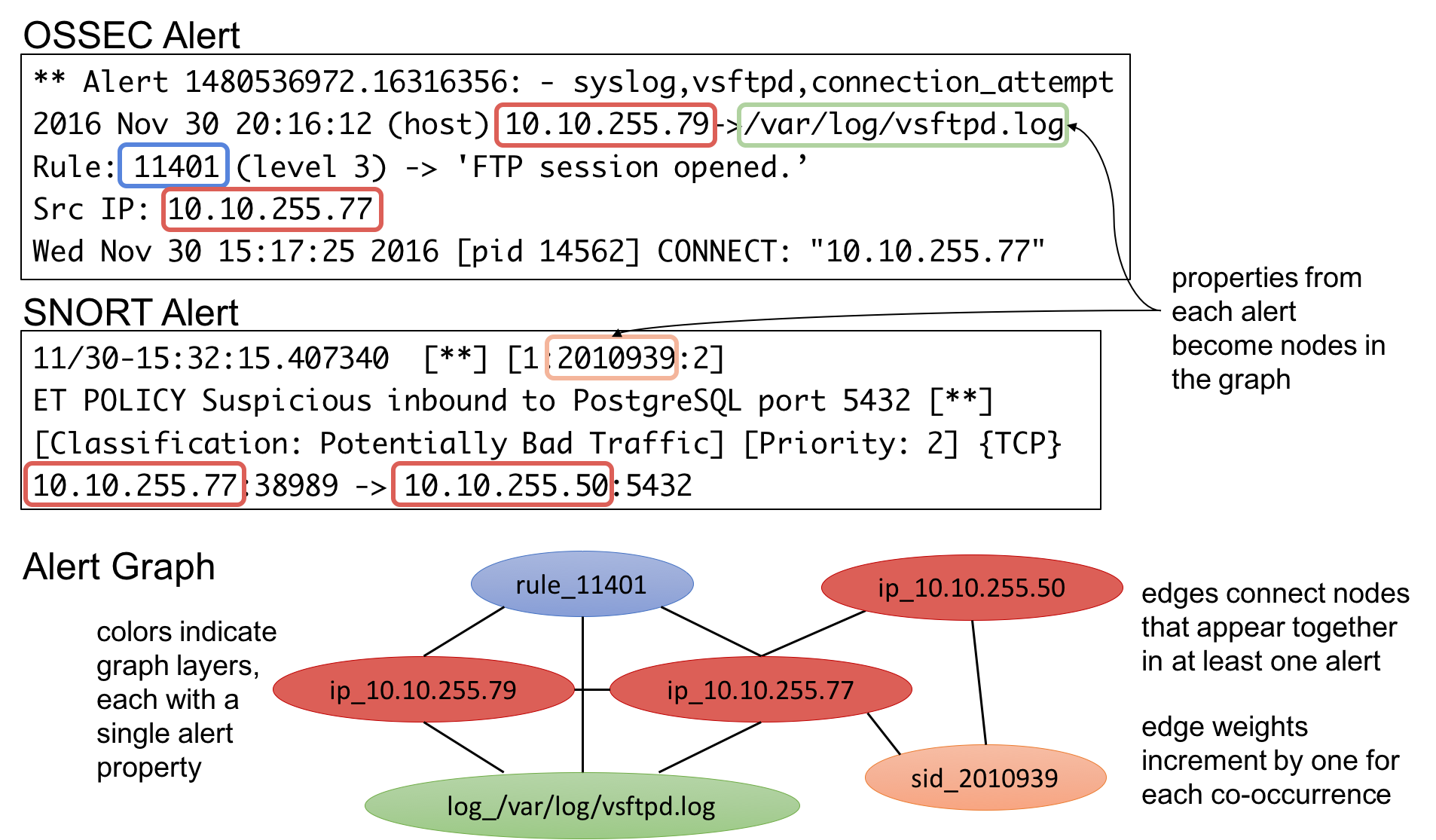}
\caption{Example of an artifact graph produced by fusing one Snort alert and one OSSEC alert.}
\label{fig:alert_graph}
\end{figure}

IDS alerts are typically formatted as structured text where the fields are populated by variables related to the triggering event (Figure~\ref{fig:alert_graph}). We organize the fields as layers in the graph (e.g., an IP address layer), with nodes that represent unique instances of the variables (e.g., a specific IP address). Edges connect nodes that appear in the same alert, incrementing in weight for each additional co-occurrence. The pseudocode in Algorithm~\ref{alg:artifact_fusion} summarizes our multi-modal data fusion (graph building) algorithm.

Every artifact type includes fields that vary in relevancy to detecting cyber artifacts. Here we use a subset of alert fields that provide orthogonal information. The classification message for a Snort alert, for example, maps to a specific signature ID. As the message can be reconstructed from the ID, the classification message does not provide new information and is excluded. For Snort alerts, we use the signature ID and the source and destination IP addresses. For OSSEC alerts, we use the source IP address (when available) and convert the hostname fields to their corresponding IP address. We also include the rule ID and the log file that generated the alert.

\begin{figure}[!ht]
\centering
\subfloat[Normal]{\includegraphics[width=0.25\columnwidth]{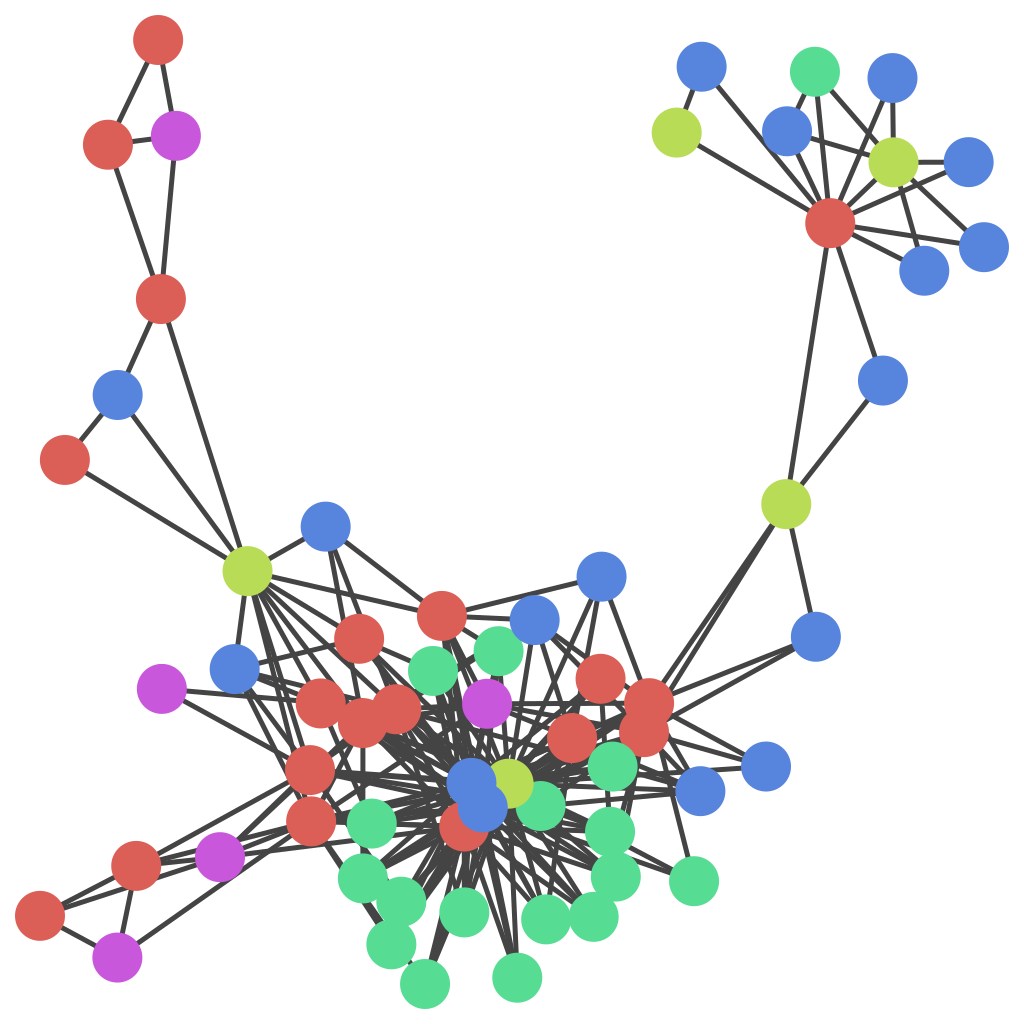}%
\label{fig_first_case}}
\hfil
\subfloat[Normal]{\includegraphics[width=0.25\columnwidth]{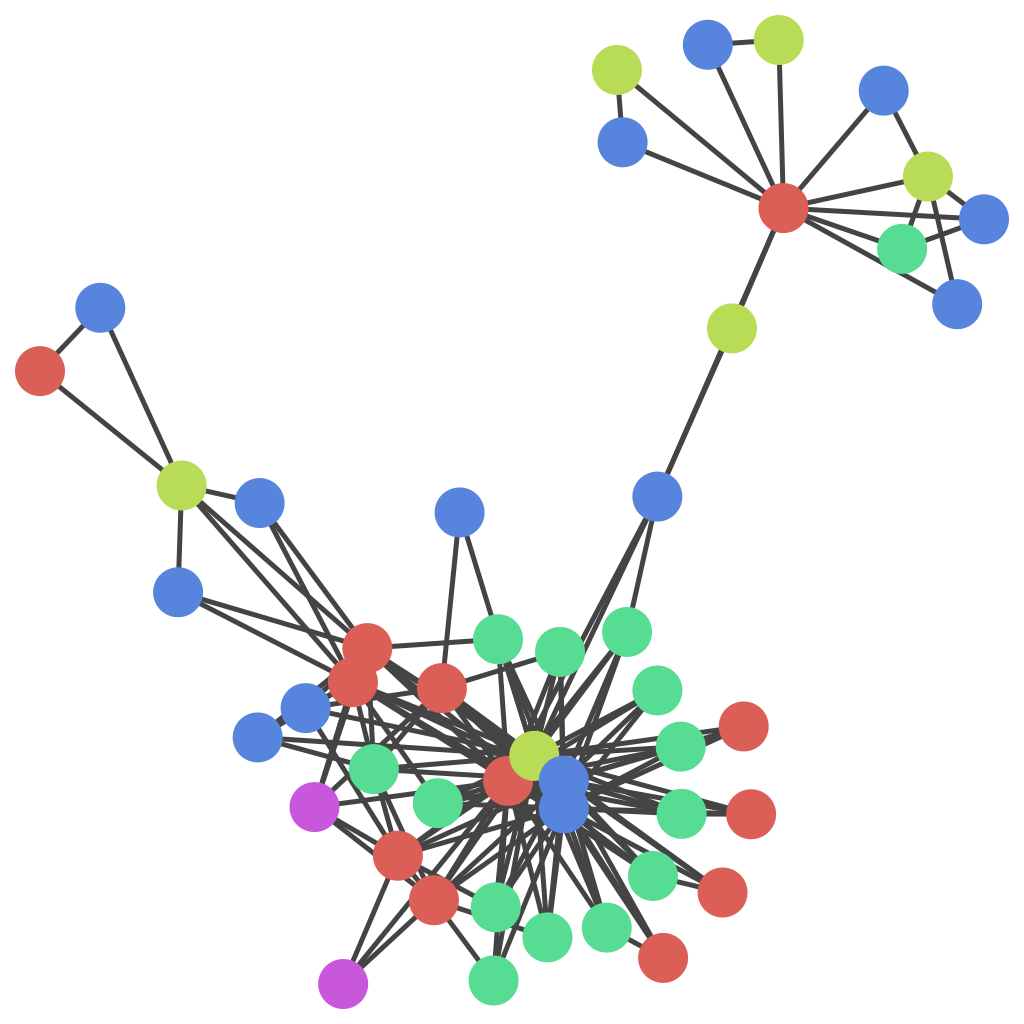}%
\label{fig_second_case}}
\hfil
\subfloat[Attack]{\includegraphics[width=0.25\columnwidth]{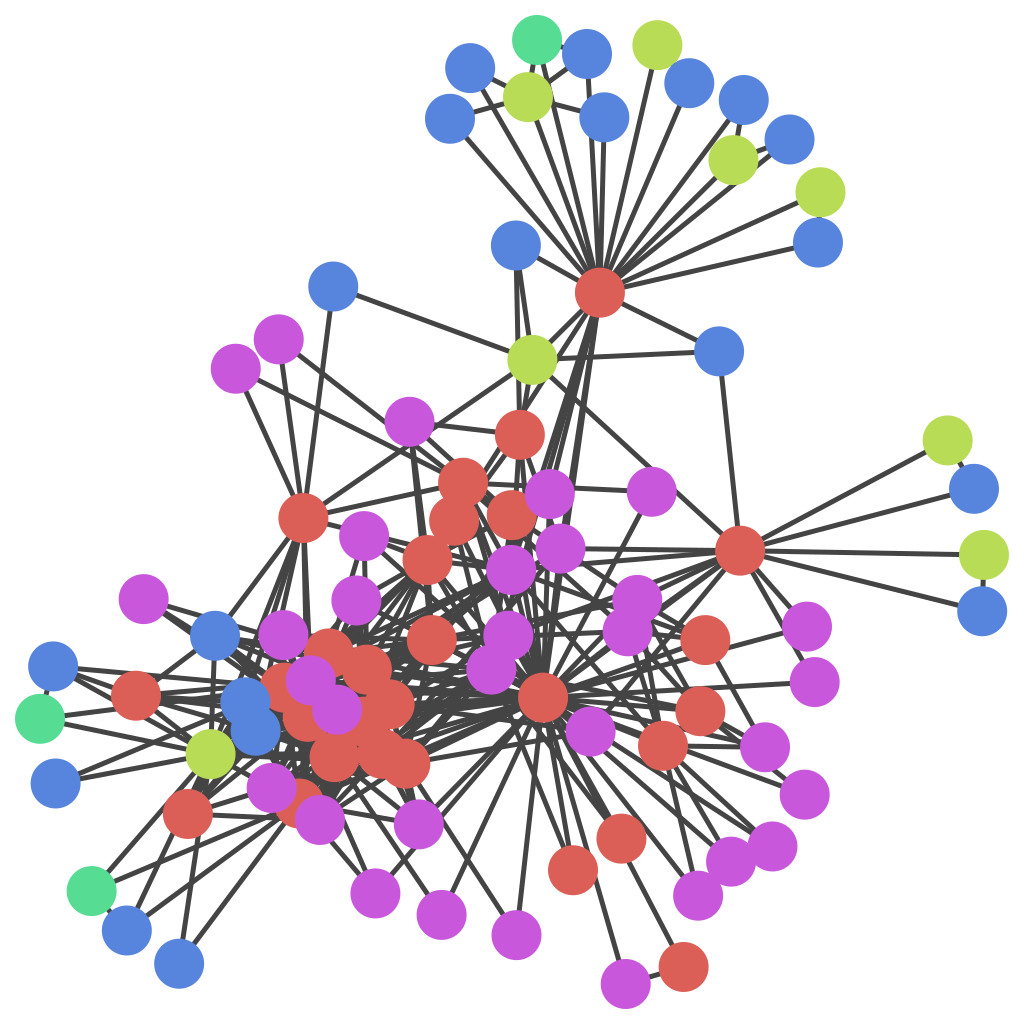}%
\label{fig_third_case}}
\caption{Comparison of artifact graphs during normal network activity and when under attack.}
\label{fig:normal_vs_attack_graphs}
\end{figure}

A graph-based approach has several advantages for detecting cyber attacks from artifacts:
\begin{itemize}
\item \emph{Context from alert interdependencies}. IDSs generate many different types of alerts (e.g., suspicious packets, failed logins). Although they may indicate different activities, alerts are not independent. For example, both alerts in Figure~\ref{fig:alert_graph} reference IP address 10.10.255.77. The graph structure explicitly links these alerts through their shared IP address node. Each alert is thus considered in the context of the other alerts in the network, which can help identify anomalies in the alert data. An alert signaling a failed login attempt, for example, may not itself be unusual. But a failed login attempt accompanied by other linked alerts in the graph may be significant. 
\item \emph{IDS fusion and efficient representation}. The graph fuses together alerts generated by disparate IDSs. Figure~\ref{fig:alert_graph}, for example, shows the fusion of alerts from a network IDS (Snort) and a host IDS (OSSEC). Other IDSs that generate alerts with overlapping fields are straightforward to include in the graph. By representing fields as nodes, we can efficiently represent the alert structure. Each panel in Figure~\ref{fig:normal_vs_attack_graphs} represents a time window with over 3,000 alerts in a single, compact graph with about 50~nodes and 200~weighted edges.
\item \emph{Robust to circumvention}. Akoglu et al.~\cite{Akoglu:2015:GBA:2757532.2757629} argue that graph-based detectors are especially difficult for adversaries to circumvent. Due to the interconnected nature of the artifact graph, an adversary would need a global view of the normal network operations (that is, the behavior of all alerts) to evade the anomaly detector. As a result, graph-based methods are widely used in fraud detection and may prove especially difficult for cyber attackers to circumvent. 
\end{itemize}

\begin{algorithm}[!ht]
    \caption{\label{alg:artifact_fusion} Build graph from IDS alerts}
    \SetKwInOut{Input}{Input}
    \SetKwInOut{Output}{Output}
    \SetKw{Continue}{continue}
    \Input{$Alerts$, a set of IDS alerts. Each $alert \in Alerts$ is parsed into key-value pairs, where each key maps to a field in the alert (e.g., rule number) and the corresponding value is the alert's value for that field (e.g., 215)}
    \Output{$G=\left \{ V,E,D \right\}$, undirected multilayer graph where $V$ are the vertices, $E$ are the weighted edges, and $D$ are the layers. Edges are defined as $(u,v,c,d,w)$ where $u,v \in V$, $c,d \in D$ such that $c$ is the layer for $u$ and $d$ is the layer for $v$. $w$ is an integer weight for the edge.}
    
    \ForEach{$alert \in Alerts$}{
        \ForEach{$key, value \in alert$}{
            \If{$key \notin D$}{
                add $key$ as new layer $d \in D$
            }
            \If{$value \notin V$}{
                add $value$ as new vertex $v \in V$
            }
        }
        \ForEach{$key_1, value_1 \in alert$}{
            \ForEach{$key_2, value_2 \in alert$}{
                \uIf{$key_1 = key_2$}{
                    \Continue
                }
                \uElseIf{$(value_1, value_2, key_1, key_2, w) \in E$}{
                    $w = w+1$
                }
                \Else{
                    add edge $(value_1, value_2, key_1, key_2, 1)$ as new edge $e \in E$
                }
            }
        }
        
    }
    \Return{$G=\left \{ V,E,D \right\}$}
\end{algorithm}

\section{Graph-Node Roles}

\begin{figure}[!ht]
\centering
\includegraphics[width=0.85\columnwidth]{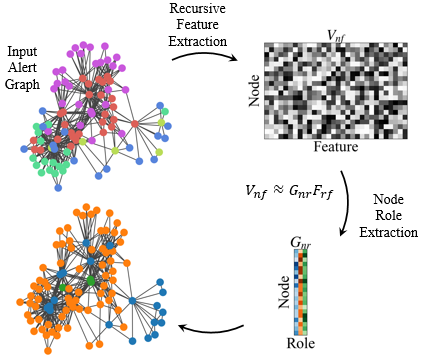}
\caption{Recursive feature extraction followed by role extraction. Alert-field nodes in the initial graph (\emph{upper-left}) are colored according to node type (signature ID, rule, IP address, etc.). Nodes in the final graph (\emph{lower-left}) are colored according to the most probable role assignment.}
\label{fig:ReFex_and_RolX}
\end{figure}

Anomalies in the artifact graph are defined as unusual changes (over time) in the graph features. Graphs contain many diverse features (e.g., degree, page-rank) but the features that are most sensitive to a cyber attack are not known in advance and may change over time. Instead of manual selection, we automatically determine the salient features and use machine learning to identify anomalies from this expanded feature set (Figure~\ref{fig:ReFex_and_RolX}). 

\subsection{Recursive Feature Extraction}

We use recursive feature extraction (ReFex)~\cite{Henderson:2011:YKG:2020408.2020512} to automatically generate feature vectors for every node in the artifact graph because it scales well and has proven successful in other applications (e.g.,~\cite{momtazpour:2015}). These feature vectors form the matrix,  $V_{nf}$, where $n$ are the nodes and $f$ are the features.
Recursive features are calculated over the node's neighbors, such as the mean of neighbor node degrees. 
Recursion continues until new recursive features are approximately linearly dependent on prior features. Here, we use both the sums and the means of neighbor node features.

The ReFex algorithm begins the recursive feature extraction with a set of primary graph features for each node:
(\emph{i}) the number of connected nodes (degree), (\emph{ii}) the number of edges connecting its neighbors (ego-network interconnectivity), and (\emph{iii}) the number of edges connecting its neighbors to other parts of the graph (ego-network out-degree). In the IDS artifact graph, edges are weighted by the number of alerts containing the connection, so we substitute weighted degree for degree to increase sensitivity to alert frequency. We also add transitivity as a fourth feature to quantify the connections between the node's neighbors. For example, if an IP address node is connected to both a rule node and a second IP address, the node's transitivity increases if the rule is also connected to the second IP address node. This might occur, for example, if both IPs were victim to the same intrusion attempt.

\subsection{Dimensionality Reduction via Role Extraction}

Anomaly detection is difficult in the high-dimensional space created by the expanded feature set. Rather than model every feature of every node for anomalies, it is convenient to summarize and model a reduced feature set. We use the role extraction algorithm (RolX~\cite{henderson:2012}) for dimensionality reduction. RolX is an unsupervised, scalable (linear in the number of graph edges) soft-clustering algorithm that reduces the features to a small set of roles.
In RolX, the node-feature matrix is factorized as 
\begin{equation}
V_{nf} \simeq G_{nr} F_{rf},
\label{eq:nf_matrix}
\end{equation}
where $G_{nr}$ is a node-role matrix in which each row quantifies the membership of a node in each role, $r$, and $F_{rf}$ is a role-feature matrix where each row defines a role in terms of the features. Non-Negative Matrix Factorization ensures role membership and feature importance are always positive~\cite{Rossi:2012:RFM:2187980.2188234}, as negative values are difficult to interpret.

\subsection{Role Number Optimization}

Evaluating~(\ref{eq:nf_matrix}) requires setting the number of roles: too few and changes in the graph may not result in role changes (increasing false negatives); too many and a node may fluctuate between two nearly identical roles (increasing false positives). The optimal number of nodes can be determined automatically~\cite{henderson:2012} using the Minimum Description Length criterion (MDL)~\cite{Rissanen:1978:PMS:2233043.2233262}.

The application of MDL is based on the insight that roles compress the node-feature matrix. Recall that the factorization in (1) is approximate; as the number of roles increases, the factorization accuracy increases but so does the model complexity. MDL balances this trade-off between accuracy and complexity by minimizing the description length $\mathcal{L} = \mathcal{M} + \mathcal{E}$, 
where $\mathcal{M}$ is the model description cost and $\mathcal{E}$ is the error cost. $\mathcal{M}$ is simply equal to $N_{b}N_{r}(N_{n}+N_{f})$ where $N_{b}$ is the number of bits per value, and $N_n$, $N_f$, and $N_r$, are the numbers of nodes, features, and roles. Note that $N_{r}(N_{n}+N_{f})$ is the total number of entries in $G_{nr}$ and $F_{rf}$. 
The error cost, $\mathcal{E}$, is calculated using the Kullback-Leibler divergence between the node-feature matrix and the factorization, 
\begin{equation}
\mathcal{E} = \Sigma_{i,j} \left( V_{i,j} \log\left(V_{i,j}/(G^\prime F^\prime)_{i,j} \right)-V_{i,j}+(G^\prime F^\prime)_{i,j} \right), 
\end{equation}
where the primes indicate the matrices are encoded using $N_{b}$ bits. 

\begin{figure}[!ht]
\centering
\includegraphics[width=\columnwidth,trim={0 0.0cm 0.01cm 0.0cm},clip]{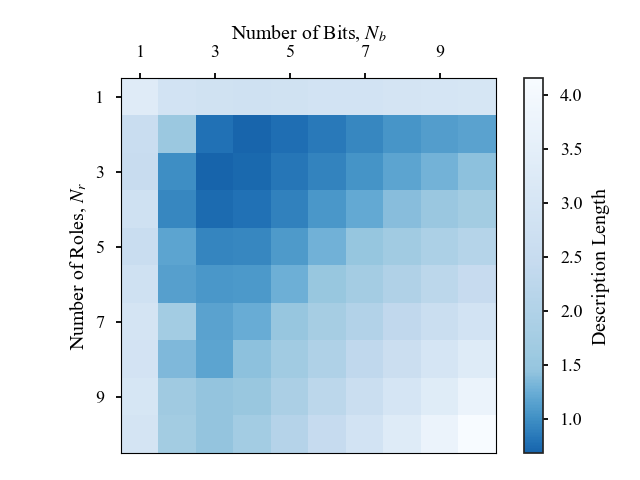}
\caption{Optimization of the number of roles finds minimum description length when using 3~bits and 3~roles.}
\label{fig:role_number_optimization}
\end{figure}

$N_{b}$ and $N_{r}$ are generally small integers, so a simple grid search reveals the minimum $\mathcal{L}$.
Figure~\ref{fig:role_number_optimization} shows the results from a graph constructed from the first seven days of the Hurricane-Panda-like APT scenario (described in Section~\ref{sec:sim_HP}). Each box in the grid is colored according to the description length $\mathcal{L}$ for that pair of $N_b$ bits and $N_r$ roles. The optimal number of roles for this graph is three.

\subsection{Role Descriptions}

A key advantage of RolX is interpretability; roles can be described in terms of intuitive features such as degree, betweenness, or pagerank. The features reflect the structural behavior of the nodes (e.g., periphery-nodes, clique-members), and are complementary to communities~\cite{FORTUNATO201075}. Compact, feature-based descriptions are especially helpful for understanding large graphs that cannot be easily visualized.

Structural descriptions may be useful to investigate:
\begin{itemize}
\item the features that define a specific role (e.g., Role~3 is defined by high betweenness values),
\item why a node has been assigned a particular role (e.g., IP address node 10.10.255.69 was assigned Role~2 because of its high transitivity and eccentricity properties), or
\item why a node has changed roles (e.g., rule node 20005 switched from Role~1 to Role~3 during the latest time step due to a large increase in betweenness).
\end{itemize}

\begin{figure}[!ht]
\centering
\includegraphics[width=\columnwidth,trim={0 0.7cm 0 0.5cm},clip]{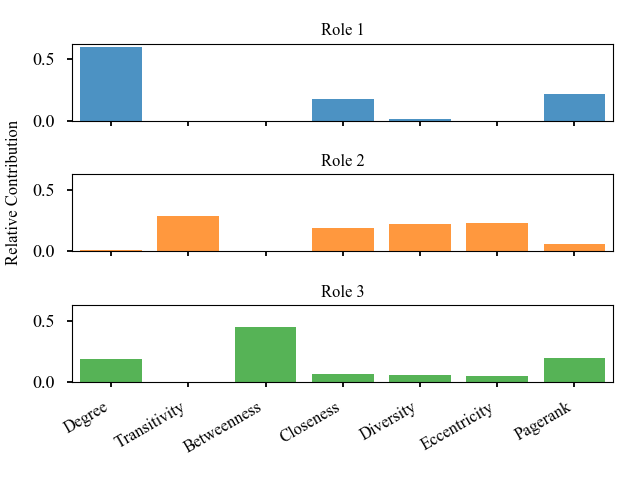}
\caption{Understanding the automatic role definitions using various graph properties. In our Hurricane-Panda-like scenario (Section~\ref{sec:sim_HP}), alert-field nodes assigned to Role 1 have high degree and pagerank, nodes assigned to Role 2 have high transitivity, diversity, and eccentricity, while nodes belonging to Role 3 have the highest betweenness.}
\label{fig:sense_making}
\end{figure}

Role descriptions are calculated by finding the non-negative matrix, $E_{rp}$, where each role, $r$, is described in terms of node properties, $p$, (e.g., degree, betweenness, etc.)~\cite{henderson:2012}, 
\begin{equation}
G_{nr} E_{rp} = M_{np} \,\,,
\end{equation}
where $M_{np}$ is a matrix of the properties for each node, and is calculated directly from the graph. The contribution of each property is normalized relative to the other properties, and to its contribution in the case of a single role, $E_{rp}/E_{rp}^\prime$, where the prime indicates the matrix was calculated using a single role. Note that these properties, $p$, are distinct from the recursive features, $f$, used in the role definitions; properties are not recursive and are manually chosen to aid comprehension.
The role descriptions for the Hurricane-Panda-like scenario are shown in Figure~\ref{fig:sense_making}.

\section{Anomaly Detection -- Graph-Node Role-Dynamics}

To identify anomalies, we divide the IDS alerts into a set of time windows, $t$, and use a training period to define a fixed role-feature matrix $F_{rf}$. We bin the remaining IDS alerts into sequential time windows, executing the following steps for each window:

  Step 1: Collect and fuse artifacts to populate the artifact graph for the current window. 
  Step 2: Extract graph features to create a node-feature matrix $V_{nf}(t)$ (e.g., ReFex~\cite{Henderson:2011:YKG:2020408.2020512}). 
  Step 3: Summarize features using RolX~\cite{henderson:2012}, assigning each node a distribution over the roles. 
  Step 4: Calculate the role distribution for each node as a function of time. Together, Steps 2--4 comprise role-dynamics described in~\cite{Rossi:2012:RFM:2187980.2188234} and below. 
  Step 5: Analyze the probabilistic role assignments over time to identify anomalous changes in the roles.
  Step 6: Investigate the anomaly cause.

To ensure that the role definitions are consistent across time steps, we modify (1) to calculate role distributions using the fixed role definitions, $F_{rf}$,
\begin{equation}
V_{nf}(t) F_{rf}^{-1} \simeq G_{nr}(t) F_{rf} F_{rf}^{-1}= G_{nr}(t)
\end{equation}
$F_{rf}$ is initially defined during the training period but can be updated periodically to adapt role definitions to changes in normalcy patterns. 

$G_{nr}(t)$ encodes a time-series for each node-role pair. There are many ways to analyze time series data for anomalies; here we analyze the average role change across all node-role pairs, 
\begin{equation}
\overline{\Delta r}(t) = \Sigma_{n=1}^{N_n} \lvert P_{n}(t) - P_{n}(t-1) \rvert / N_n , 
\end{equation}
where $P_{n}(t)$ is the maximum role membership probability for node $n$ in time window $t$. Nodes will not necessarily appear in every time window, and $P_{n}(t)$ is set to null until its first occurance. Once a node appears, its previous probability of role membership fills forward into new time bins where $n \notin V$. For the scenarios here, setting a constant threshold on $\overline{\Delta r}(t)$ is sufficient to identify anomalies corresponding to the start of the attacks.

\section{Test and Evaluation}

\subsection{Simulation}

\begin{figure}[!ht]
\centering
\includegraphics[width=\columnwidth,trim={2.8cm 8.8cm 2.7cm 4.0cm},clip]{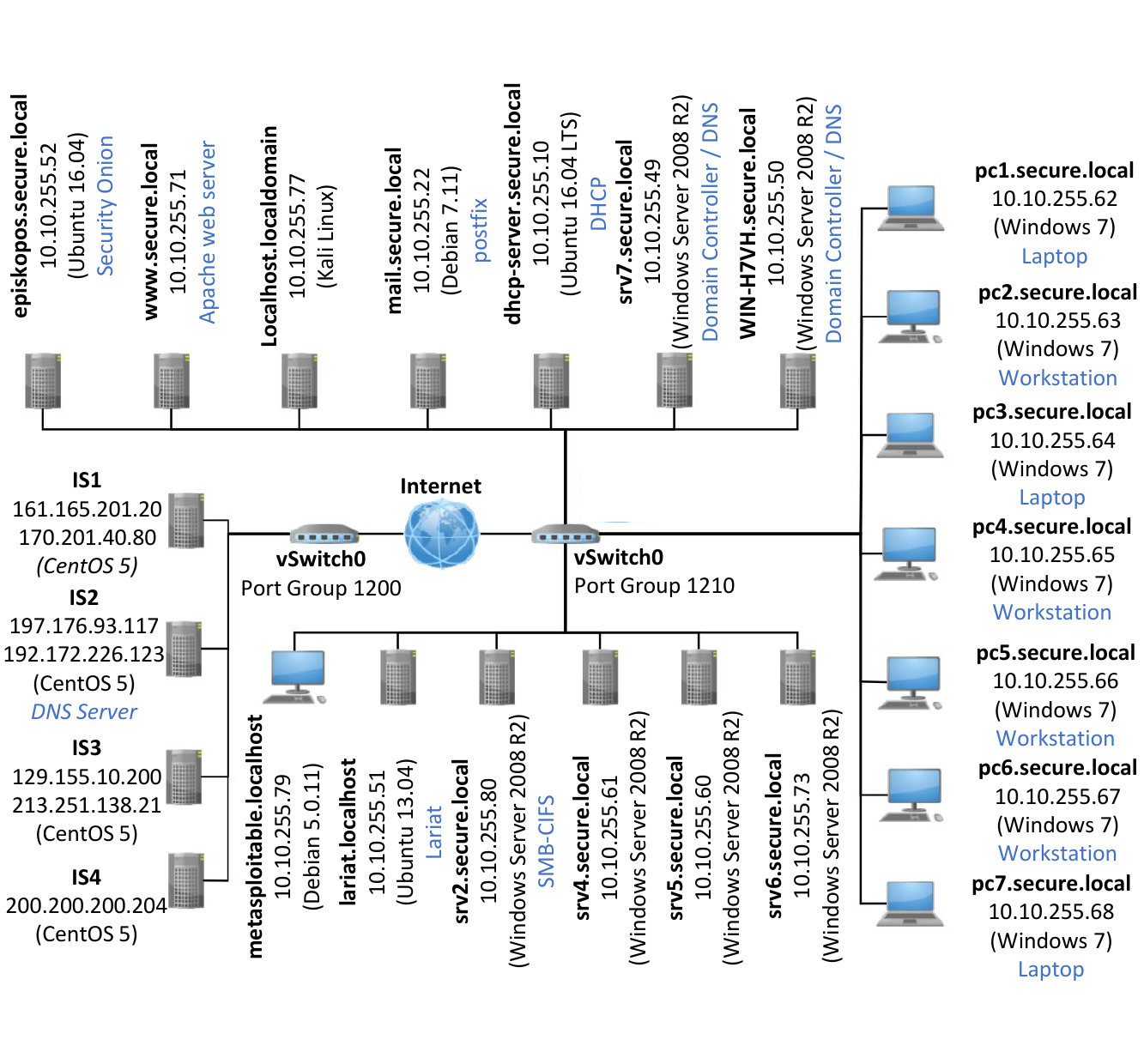}
\caption{The simulated network consists of 24~virtual machines (Windows and Linux workstations, as well as data, print, and DNS servers), virtual switches, and a virtual server approximating connections to the internet.}
\label{fig:simulated_network}
\end{figure}

To evaluate the ability of our approach to identify anomalies in IDS alerts, we use IDS alerts generated from a network of virtual machines (Windows and Linux workstations, data servers, print servers, and DNS servers), virtual switches, and a virtual server that approximates connections to the internet (Figure~\ref{fig:simulated_network}).
The network includes installations of Snort and OSSEC 
which generated alerts throughout the  experiment. 

The simulations include three weeks of normal network operations, background traffic (packet flow), and IDS alerts, as well as two distinct APT-like cyber attacks that occur near the end of each simulation.
We used the first week of the simulation (all prior to the attack) to obtain the initial role definitions, $F_{rf}$, and the optimal number of roles. 
The remaining alerts are divided into non-overlapping 8-hour windows, and the alerts in each window populate an artifact graph.
$G_{nr}(t)$ is calculated using Steps 1--4 in Section~IV and $\overline{\Delta r}$ is calculated using (5). Any values of $\overline{\Delta r}$ above the preset threshold are flagged as anomalous.

\subsection{Simulated Attack 1}\label{sec:sim_HP}

The first scenario is characteristic of Hurricane Panda, an APT that targets infrastructure companies and is thought to be of Chinese origin~\cite{lemay:2018}.
The simulation encompasses network operations from from 9~Nov 2016 to 3~Dec 2016, with the first attack command being issued on 30 Nov at 8~PM (UTC), and the final attack command occurring on 2 Dec at 10:03~PM (UTC). The attacker in our simulation used database injection to steal authentication information and gain initial access to the network. After waiting 24~hours, the attacker used this access to steal additional authentication information for other users. Using the credentials stolen from another user, the attacker moved laterally through the network, installing post-exploitation tools (mimikatz\footnote{https://www.offensive-security.com/metasploit-unleashed/mimikatz/}) on several windows machines. Waiting another 24 hours, the attacker then used the post-exploitation tools to disable the firewalls and task schedulers for several windows machines on the network.

\begin{figure}[!ht]
\centering
\includegraphics[width=\columnwidth,trim={0 3.5cm 0 0.0cm},clip]{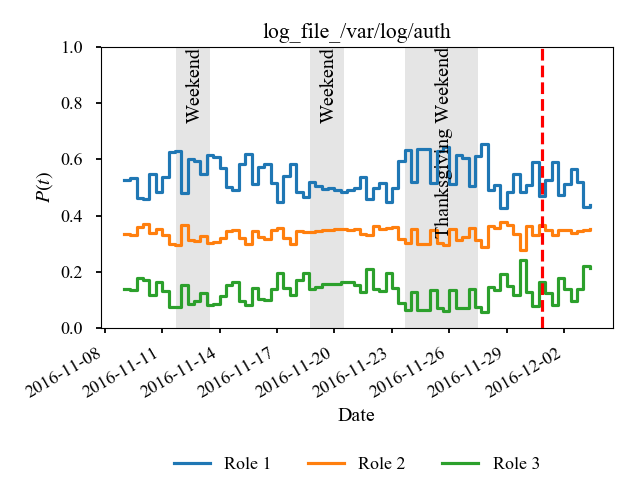}
\includegraphics[width=\columnwidth,trim={0 3.5cm 0 0},clip]{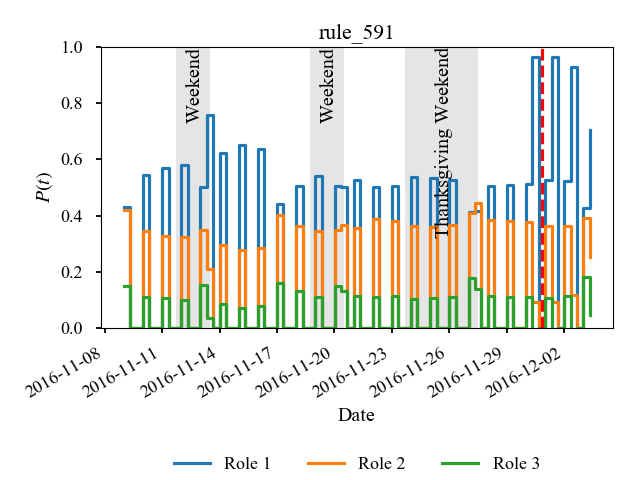}
\includegraphics[width=\columnwidth]{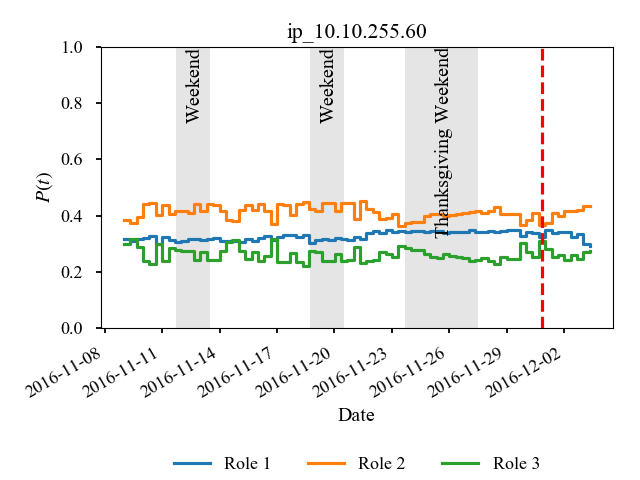}
\caption{Probability of role membership for alert-field nodes in the artifact graph corresponding to a log file node (\emph{top}), an OSSEC rule node (\emph{middle}), and an IP address node (\emph{bottom}). The vertical dashed line indicates the start of Simulated Attack~1.}
\label{fig:logfile_membership}
\end{figure}

The entire set of 364,080~IDS alerts can be stored as 263~nodes with 1625~weighted edges in our alert graph representation, illustrating the scalability of our approach. From these alert graphs, ReFex identified 112 recursive features based on the weighted node degree, ego-network interconnectivity, ego-network out-degree, and transitivity features. The optimal number of roles is three (Figure~\ref{fig:role_number_optimization}). Role descriptions are shown in Figure~\ref{fig:sense_making}.
Figure~\ref{fig:logfile_membership} shows the role membership over time for three typical alert-field nodes in the graph; a log file node, an alert rule node, and an IP address node. 

\begin{figure}[!ht]
\centering
\includegraphics[width=\columnwidth,trim={0 3.2cm 0 0.3cm},clip]{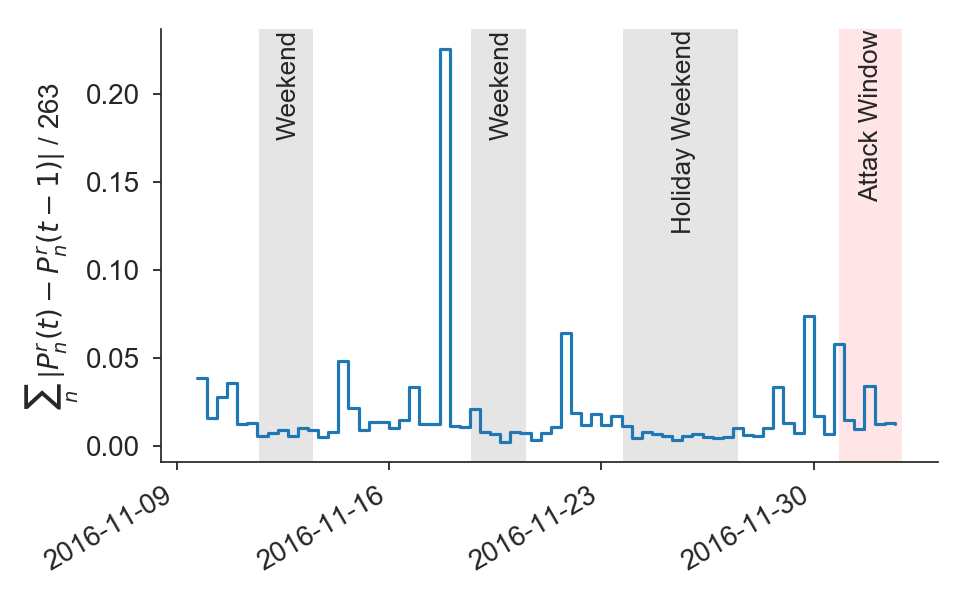}
\includegraphics[width=\columnwidth,trim={0 3.2cm 0 0.0cm},clip]{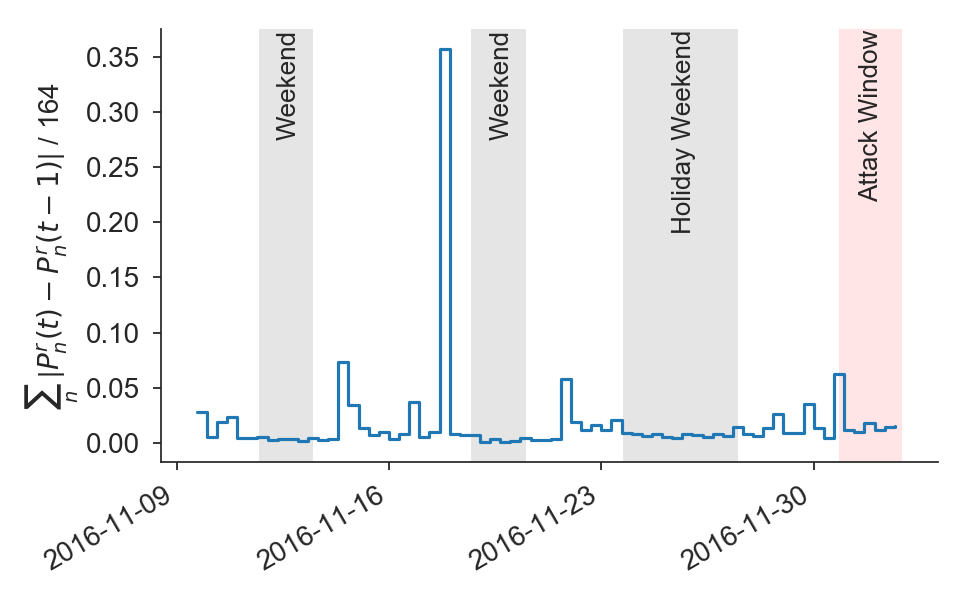}
\includegraphics[width=\columnwidth,trim={0 0.0cm 0 0.0cm},clip]{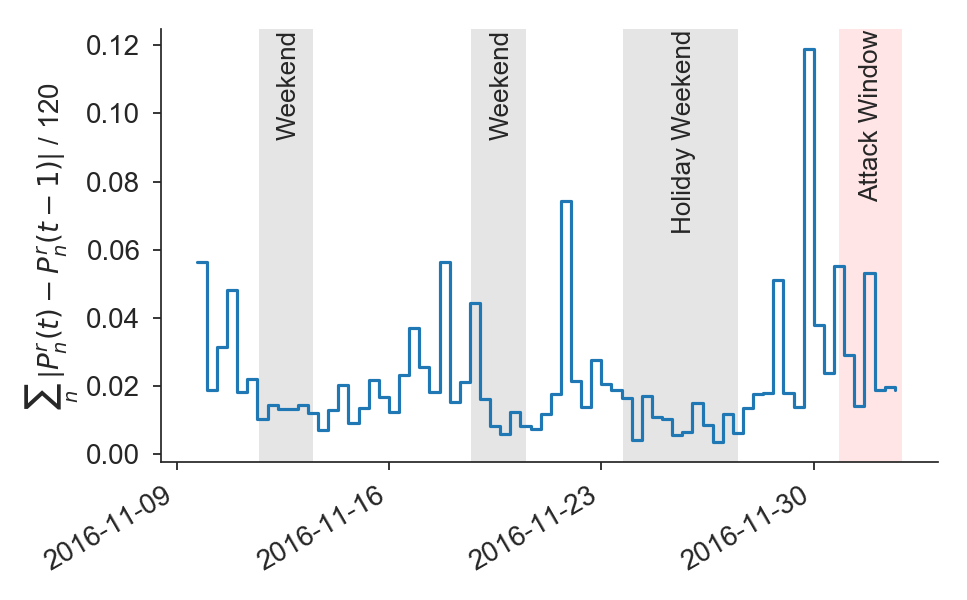}
\caption{Anomaly detection using role dynamics in the Hurricane-Panda-like attack simulation from fused Snort and OSSEC alerts (\emph{top}), only Snort alerts (\emph{middle}), and only OSSEC alerts (\emph{bottom}). The shaded region on the right indicates the simulated attack window.}
\label{fig:IDS_independence}
\end{figure}

Using a manually-defined threshold of 0.05, our approach identifies four anomalies in $\overline{\Delta r}(t)$ (Figure~\ref{fig:IDS_independence}). The first two are related to IP address changes (10.10.255.40 appears on 15~Nov, and 10.10.255.49 appears on 18~Nov). The fourth correlates with the start of the attack (the shaded attack region in Figure~\ref{fig:IDS_independence}). 
Of the 364,080~IDS alerts (Snort and OSSEC) produced by the simulation, only 316 are related to the attack, yet our unsupervised machine learning approach identified the start of the attack as anomalous. 
The  anomalies also appear when restricting the data to include only Snort or OSSEC alerts, suggesting the approach may be robust to the specific IDSs available on the network.

\subsection{Simulated Attack 2}

The second scenario is characteristic of Energetic Bear (a.k.a. Crouching Yeti, Dragonfly, or Havex), an APT that has targeted defense and aviation companies in the U.S. and Canada, and energy firms in Europe~\cite{lemay:2018}. Artifacts in the code suggest Russian-speaking authors, but the origin remains uncertain. In this simulated scenario, the attacker used an email-based phishing attack to direct a network user to a malicious webpage. While on the webpage, that user's browser downloaded an exploit that gave the attacker access to the machine. Now able to access the network, the attacker installed post-exploitation tools (mimikatz) to establish a persistent foothold and steal authentication information of other users. The attacker then moved laterally throughout the system, using a remote desktop exploit to connect to other machines and create new administrator users on those machines. Having established a way to maintain access and control of these machines, the attacker cleaned up logs and other traces of the attack. 

This second scenario encompasses network operations from 1~Jan 2017 to 4~Feb 2017. The attack occurred on 31~Jan between 4:57~PM (UTC) and 7:00~PM (UTC). OSSEC generated 702,241 IDS alerts throughout this entire period, which can be represented as 90~nodes and 475~weighted edges in our artifact graph representation.
As with the Hurricane Panda scenario, the first week of data was used as a normalcy period to automatically determine the initial optimal number of nodes, the recursive features, and the role definitions. 

\begin{figure}[!ht]
\centering
\includegraphics[width=\columnwidth,trim={0cm 3.2cm 0cm 0.0cm},clip]{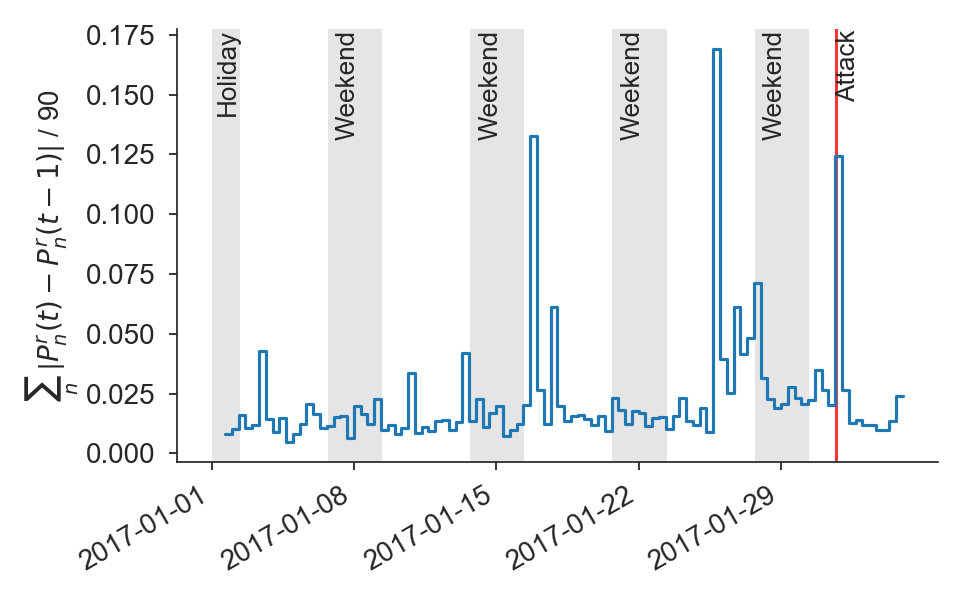}
\includegraphics[width=\columnwidth,trim={0cm 0.0cm 0cm 0.0cm},clip]{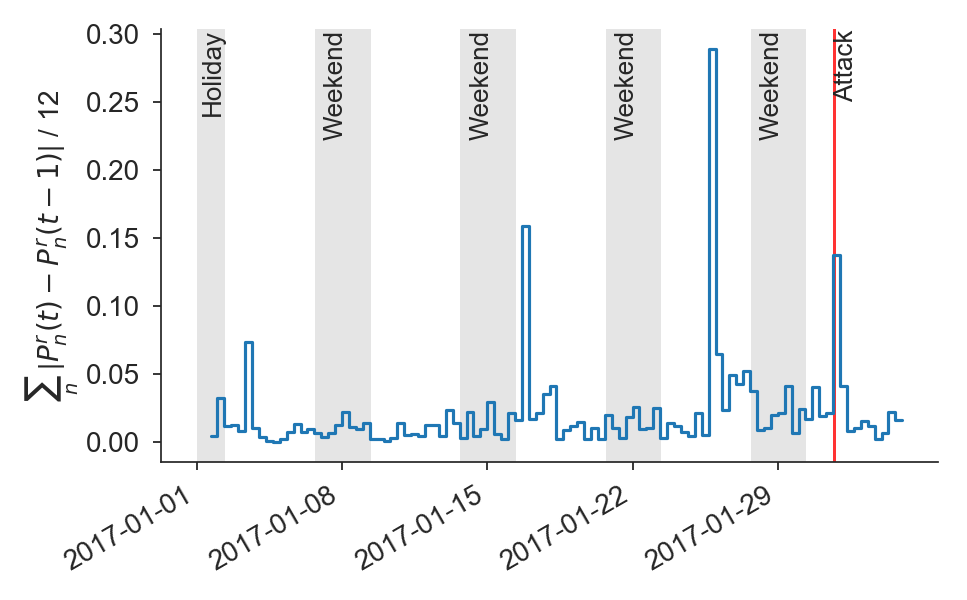}
\includegraphics[width=\columnwidth,trim={0cm 3.2cm 0cm 0.3cm},clip]{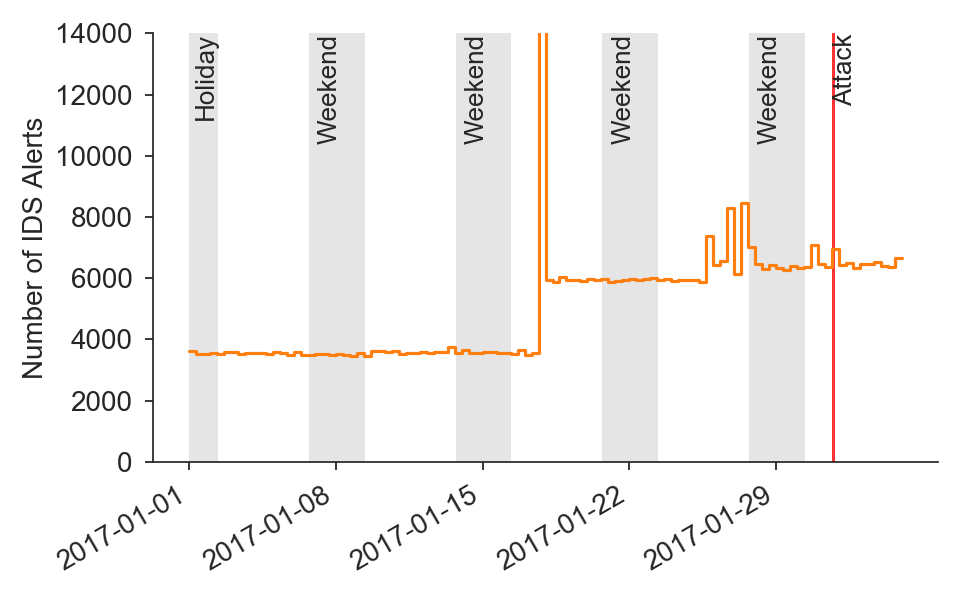}
\caption{Anomaly detection using role dynamics in the Energetic-Bear-like attack simulation from OSSEC alerts showing the incremental change in probabilistic role assignment normalized over all 90 alert-field nodes in the artifact graphs (\emph{top}) and only considering the alert-field nodes in the ``log file'' layer (\emph{middle}).  The vertical line to the right indicates the time bin during which the simulated attack occurred. Total number of IDS alerts throughout the simulation (\emph{bottom}). }
\label{fig:EB_role_dynamics}
\end{figure}

Using a manually-defined threshold of 0.1, we identified three anomalies (Figure~\ref{fig:EB_role_dynamics}). We found the same three anomalies when considering all alert-field nodes (Figure~\ref{fig:EB_role_dynamics}, \emph{top}) and when restricting the analysis to one layer, e.g., only ``log file'' nodes (Figure~\ref{fig:EB_role_dynamics}, \emph{middle}). The third anomaly coincides precisely with the simulated Energetic-Bear-like attack campaign. This promising result suggests that this formalism for anomaly detection may robustly identify anomalous behavior from different types of attacks (e.g., Hurricane-Panda-like, Energetic-Bear-like, etc.).

Note that this attack-related anomaly does not correlate with an unusual change in the number of alerts (Figure~\ref{fig:EB_role_dynamics}, \emph{bottom}), as the method is sensitive to alert interconnections (which affect the nodes and edges of the alert graph), rather than simply alert volume (which affects the edge weights only). Contrast this with the lack of an anomaly over the 0.1~threshold during the large spike in alert volume near the middle of the dataset. Despite the large increase in alert volume (perhaps comprising many identical alerts), the alert graph does not change enough to cause anomalous role dynamics. Our graph-based anomaly detector provides information complementary to alert volume analyses; in this example the attack/anomaly is much easier to find using role dynamics than volume analyses alone.

\section{Discussion}

With any model that learns normal behavior from real training data, there is always some concern that the training data includes an  attack. Including the attack in the definition of normal prevents identifying those attacks as anomalous, increasing false negatives. This is a real risk, but there are some mitigating factors that are useful to consider. 

The first mitigating factor is the frequency of malicious activity. If the attacker has gained access to the network but is active only infrequently to disperse any indicators of compromise, the probability of training on attack-related artifacts is small. If attack components are included in the training data but are missing from later time windows, this is also an anomaly that could be identified by our method. Comparing results from multiple training periods may also provide insight; large unexpected changes unrelated to known network operations should be investigated. 

The second mitigating factor is escalation. Artifacts from malicious events included in the training data will likely not encompass all phases of the cyber kill chain. Escalating the attack to new phases should generate new cyber artifacts. As these new artifacts are not included in the training data, they would be identified as anomalous. 

The third mitigating factor is the ability to detect changes in the attack methods. Advanced attackers, such as APTs, build custom tools and frequently adapt their methods~\cite{lemay:2018}. If some of the methods are included in the training data, only those methods will be learned as normal. New attacks not included in the training data would be identified as anomalous. The ability to detect new, previously unseen attack steps is an advantage of our unsupervised approach.

Although the role-dynamics method includes an automatic method for determining the optimal number of nodes, there remain several hyperparameters that must be defined. These include which types of cyber artifacts to use (e.g., IDS products), which artifact fields (e.g., IDS alert fields) to populate the graph, the length of the training period, the time window size, and the anomaly threshold. In the examples considered here, these choices were dictated by domain knowledge about the IDS alerts and the limitations of the dataset. As each network configuration differs, using detailed knowledge about the network may be the best approach for determining the hyperparameters. It may also be possible to optimize these hyperparameters to increase sensitivity to cyber attacks and decrease sensitivity to benign anomalies using either known test cases, or prior attacks on the network. We leave this for future work. 

\section{Conclusions}
Graph-based anomaly detection is a promising new approach for detecting cyber attacks from cyber artifacts. In both test scenarios, our method identified anomalies corresponding to the start of the attack.
We fused data from several intrusion detection systems into artifact graphs using a novel graph construction based on fields in the triggered alerts, not the network topology. Analyzing the role dynamics in these graphs for simulated Hurricane-Panda-like and Energetic-Bear-like APT datasets, we identified a handful of anomalies, including anomalies that coincide with the start of each  attack. Our approach successfully identified simulated attacks through anomalies in IDS alert patterns, and reduced the number of false-positive alerts from thousands to just three false-positive anomalies (in one simulation) and two (in the other simulation). Our results illustrate how graph-node role-dynamics analyses can identify anomalies in IDS alerts, however causal analysis will require further investigation by human analysts.


\begin{acks}
The authors would like to thank Dr. Angelos Keromytis for his support and guidance throughout this research effort, and Dr.~Ron Watro for  his helpful insights, recommendations, and direction.
The authors would also like to thank the anonymous referees for their valuable comments.
This research was supported by the Defense Advanced Research Projects Agency (DARPA) under Contract No.~W31P4Q-15-C-0069, 
and this paper has been approved with Distribution Statement ``A'' (Approved for Public Release, Distribution Unlimited).
The views, opinions, and/or findings contained in this document are those of the authors and should not be interpreted as representing the official policies, either expressed or implied, of DARPA or the U.S. Government.
\end{acks}

\bibliographystyle{ACM-Reference-Format}
\bibliography{bibliography}

\end{document}